# Challenges in designing ethical rules for Infrastructures in Internet of Vehicles

Razi Iqbal

razi.iqbal@ieee.org

*Abstract*— Vehicular Ad-hoc Networks (VANETs) have seen significant advancements in technology. Innovation in connectivity and communication has brought substantial capabilities to various components of VANETs such as vehicles, infrastructures, passengers, drivers and affiliated environmental sensors. Internet of Things (IoT) has brought the notion of Internet of Vehicles (IoV) to VANETs where each component of VANET is connected directly or indirectly to the Internet. Vehicles and infrastructures are key components of a VANET system that can greatly augment the overall experience of the network by integrating the competencies of Vehicle to Vehicle (V2V), Vehicle to Pedestrian (V2P), Vehicle to Sensor (V2S), Vehicle to Infrastructure (V2I) and Infrastructure to Infrastructure (I2I). Internet connectivity in Vehicles and Infrastructures has immensely expanded the potential of developing applications for VANETs under the broad spectrum of IoV. Advent in the use of technology in VANETs requires considerable efforts in scheming the ethical rules for autonomous systems. Currently, there is a gap in literature that focuses on the challenges involved in designing ethical rules or policies for infrastructures, sometimes referred to as Road Side Units (RSUs) for IoVs. This paper highlights the key challenges entailing the design of ethical rules for RSUs in IoV systems. Furthermore, the article also proposes major ethical principles for RSUs in IoV systems that would set foundation for modeling future IoV architectures.

*Keywords—Ethics; Road Side Units; Vehicular Ad-hoc Networks; Internet of Vehicles; Intelligent Transportation Systems*

## I. INTRODUCTION

Intelligent Transportation Systems (ITS) have been transforming the traditional transportation systems into intelligent systems for decades now. ITS have seen tremendous advancements throughout their life time and have become an essential part of transportation in developed and developing countries. VANETs have played a significant role in transmuting the notion of connected vehicles and infrastructures in ITS. Introduction of autonomous vehicles and drone RSUs are bringing dynamicity with more independence and decentralization to the transportation systems [1]. VANETs have been key area of research for both academicians and industry professionals for years that has reshaped the overall perception of ITS for concepts like smart cities.

IoV is an adherent of VANET where each unit of the system is Internet enabled. This Internet connectivity provides system with greater capabilities of sharing information of common interests besides providing more opportunities of medium of communication [2]. Recent advancements in computing and communication technologies like EDGE Computing, Grid Computing, Parallel Processing, Big Data Analysis, Web Semantics and Artificial Intelligence has opened horizons of opportunities for developing and deploying safety and infotainment applications for IoV systems.

One of the key advantages of Internet connectivity in IoV is socializing of objects of systems, e.g., vehicles, infrastructure, passengers, drivers and environmental sensors etc. Sharing of information on roads through Internet provides ease of driving, safety, awareness, warnings, traffic updates and special services like discount coupons, vacant car parking information and alternate routes etc. [3]

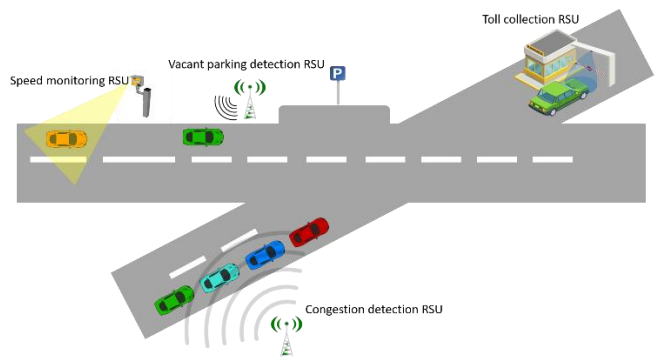

Fig. 1. Distinct types of RSUs in IoV System

Infrastructure sometimes referred to as Road Side Units (RSUs) are considered key component of IoV systems. RSUs are essential for IoV systems in a way that they provide sensing, communication, processing and computing capabilities. One of the most common RSUs these days are speed checking RSUs that are equipped with cameras for measuring the speed of vehicles on roads. Besides speed checking RSUs, other examples are toll collection RSUs, smart bus stops, digital billboards and motion detecting RSUs. Majority of the RSUs mentioned are connected directly or indirectly to the cloud (Internet) to store, process, compute and communicate information [4]. Moreover, multifaceted RSUs are also gaining popularity in a way that they are capable of multitasking and multiprocessing. Fig. 1 illustrates distinct types of RSUs in IoVs systems.

Growing autonomy in IoV systems brings the concern of ethics. IoV systems will be challenged with ethical dilemmas and anticipated to operate in an ethically responsible way. Self-governed operations of IoV systems require setting ethical rules and policies before these systems can take an autonomous decision [5]. However, currently, no ethical guidelines are available for components of IoV systems that results in lack of applications and services. This article is an effort towards setting the foundation for designing ethical rules for one of the key components of IoV systems, the RSUs. Below are the contributions of this article:

- Highlight the key challenges involved in designing ethical rules for RSUs in IoV systems
- Propose major ethical principles for RSUs in IoV systems that would set foundation for modeling future IoV architectures.

The rest of the article is organized as follows: Section II provides more details of RSUs and their types currently available in literature and their applications in real world. Section III focuses on the challenges of designing ethical principles for RSUs in IoV environment. Section IV proposes four major ethical rules for RSUs in IoV systems that are expected to set foundation for future ethical IoV architectures. Finally, the conclusion section concludes the article.

## II. ROAD SIDE UNITS IN IOV SYSTEMS

RSUs are considered one of the key components of IoV systems. RSUs perform several activities like traffic monitoring, speed checking, congestion detection, toll collection, identifying vacant car parkings, surveillance, traffic jams, warnings and safety and security on roads. Such activities fall under sensing capabilities of RSUs. In order to realize sensing, RSUs are equipped with additional integrant like cameras and environmental sensors [6].

Besides sensing, RSUs are capable of computing and processing of information as well. In IoVs, RSUs can process and compute information locally to save time and provide quick access to information on request from peer entities of the network, or it can leverage the capabilities of the cloud for data processing if local resources are not sufficient.

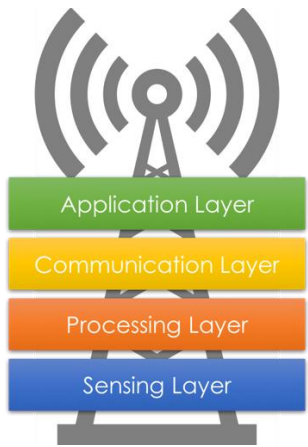

Fig. 2. Traditional RSU layered Architecture

In VANETs, RSUs normally use a dedicated communication technology, Dedicated Short Range Communication (DSRC) to communicate with peer RSUs and vehicles on roads. In IoVs, RSUs are connected directly or indirectly to the Internet. In case of indirect connection to the Internet, RSUs utilize DSRC or wired connection with other Internet enabled RSUs. Furthermore, in IoVs, RSUs can be multifaceted to incorporate multiple communication technologies like Cellular, Wi-Fi, DSRC, 6LowPAN and Wi-Max etc.

RSUs provide diverse applications based on the context. In IoVs, application layer is anticipated as the most comprehensive layer of RSU architecture as it can provide local and cloud based applications and services. Some of the applications and services provided by RSUs in IoV are warning and safety messages, information about nearby restaurants, maps for navigation within close vicinity, infotainment applications like media sharing and Internet sharing etc. Fig. 2 illustrates the traditional RSU layered architecture.

This section highlights different categories of RSUs currently available in literature along with their real world applications.

### A. Data Collection RSUs

One of the most popular RSUs these days are data collection RSUs. These are the units that are equipped with several types of sensors to collect information [7]. One of the commonly used data collection RSUs are speed monitoring RSUs that are equipped with high definition cameras that measure the speed of the vehicle based on images captured at two distinct locations. Fig. 3 illustrates the working of such RSUs.

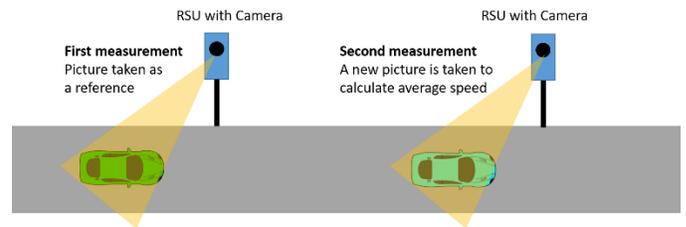

Fig. 3. Speed measuring RSUs

As illustrated in the figure, in order to measure the exact speed of the vehicle, an image is captured for the vehicle at two stages. The first image is captured as a reference and the second image is captured to calculate the average speed of the vehicle. Based on the speed calculated by the camera, fines are issued by the law enforcement agencies.

### B. Toll Collection RSUs

Another popular type of RSUs is toll collection RSUs that are normally installed at highways to collect highway charges. Current state-of-the-art technology used in such RSUs is Radio Frequency Identification (RFID) that operates through an initiator (reader) and responder (passive tag) installed at collection booth (toll gate) and vehicle respectively.

As illustrated in Fig. 4, as soon as the vehicle enters the communication range of the RFID initiator (approx.10m), an electromagnetic field is generated around the vicinity that powers up the passive tag installed in the vehicle. Based on the information stored in the tag which is bundled with the account details of the driver, charges are deducted from the account.

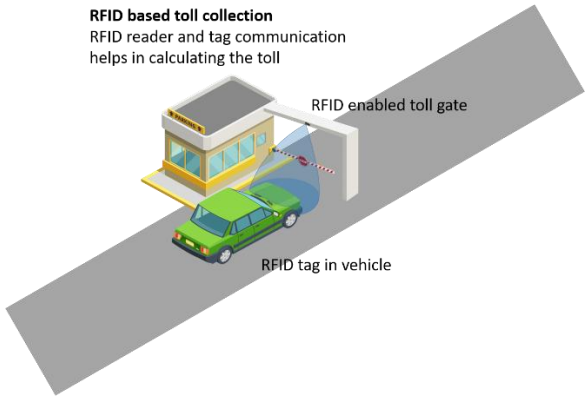

Fig. 4. Toll collecting RSUs

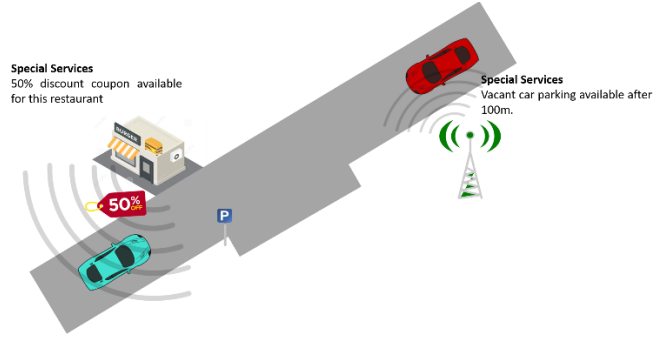

Fig. 5. Special Services RSUs

## C. Information Dissemination RSUs

A new form of RSUs emerging these days is information dissemination RSUs. These RSUs are connected to each other through wired or wireless communication medium to form a network in order to shared information [8]. As soon as an event occurs near a RSU, this information is disseminated to the other RSUs in the network based on the context. Fig. 5 highlights the working of information dissemination RSUs.

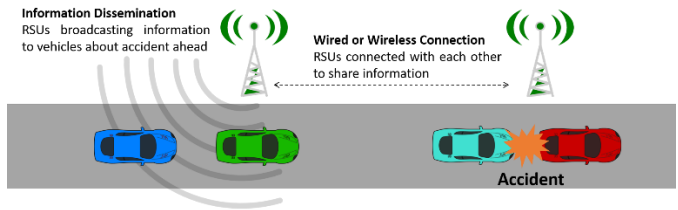

Fig. 5. Information Disseminating RSUs

As illustrated in Fig. 5, two RSUs are connected to each other through a wired of wireless connection. As soon as an accident occurs near one of the RSUs, this information is transmitted to the other RSU in the network which ultimately broadcasts this information to nearby vehicles. This information dissemination helps in avoiding traffic jams due to unfortunate incidents on roads.

## D. Special Services RSUs

Surge in Internet connectivity on roads have conceptualized several new forms of services to facilitate vehicles, drivers and passengers. RSU is one of the components of IoV system that can provide special services to other components of the system based on demand. Such RSUs are categorized as special services RSUs [9]. The special services provided by these RSUs are internet sharing, discount coupons of restaurants, vacant car parking detection, upcoming gas station information, alternate routes, media sharing, chatting groups and many other social networking services.

Fig. 6 illustrates the working of two cases of special services RSUs. In first case, a restaurant is providing a discount coupon of 50% to the passing vehicles. This information is transmitted to passing vehicles through the Internet since both restaurant (considered as a RSU since it can transmit and receive information) and vehicle are Internet enabled. Similarly, in the second case, RSU communicates with a passing vehicle to inform it about upcoming vacant car parking.

## III. CHALLENGES FOR ETHICAL RULES FOR RSUs IN IoV

RSUs are smart sensing, computing, processing, communicating and storing devices that provide required services without human interferences. However, these self-operational capabilities of RSUs bring ethical concerns when it comes to information gathering, information dissemination and decision making. Currently, no significant efforts have been made to design ethical guidelines for RSUs in IoV systems because of several challenges. This section highlights the major challenges involved in designing ethical guidelines (rules or policies) for RSUs in current state-of-the-art IoV systems.

### A. Architectural Design for RSUs

IoV is a relatively new concept, however, RSUs have been around for decades. Besides their availability for years, there is no standard architectural design available for RSUs [10]. In various parts of the world, different types of RSUs are used based on distinct architecture. Several architectural design choices are available for RSUs, e.g., RSUs with sensing and communicating layers, RSUs with sensing, processing and communicating layers and RSUs with only application layer. For example, congestion detection RSUs rely on sensing and communication layers, speed monitoring RSUs rely on sensing, processing and communicating layers and digital billboards on highways rely on application layer only.

The architecture of some RSUs is very complicated as they include security, privacy, trust management, gateway, EDGE computing and web services layers besides traditional layers presented in Fig. 2. Lack of standard architectural design encompasses the challenge of designing ethical rules for RSUs. Scheming generic ethical rules for all kinds of RSUs appears unrealistic unless a global standard for RSU architecture is defined.

### B. Decentralization and Scalability

Internet connectivity in IoV has enabled different components of the system to communicate with each other without relying on other units. The concept of decentralization holds importance as it allows each component of the system to sense, process, communicate and react to events associated with it. Independent operations of IoV components provides flexibility, context-awareness, localization and less dependability on other units. Decentralization provides heterogeneity to the system as each component of IoV system can be different from the other component depending upon the type of the component [11]. IoV system incorporates several vehicles and each vehicle can be different from its peer vehicle in make, kind, size and color etc. If RSUs have to communicate with these different vehicles, ethical rules should be defined on communication, sensing, processing and information storage at RSU level. For example, if a vehicle does not want to share its information and still RSUs are collecting vehicle information like car type, color, make, registration number plate and driver information etc., it would be an ethical concern for the driver of the vehicle. Defining an ethical rule for each vehicle, driver and passenger is extremely difficult in IoV environment as the network is very large and probability of predicting the scalability of the network is almost impossible in large area networks.

*C. Communication Technologies*

In IoV systems, information is transmitted using long range and short range wireless technologies except the communication between RSUs which is sometimes wired. However, the latest trend in RSU communication is wireless transmission as it provides flexibility, cost effectiveness, less hassle and scalable networks. Commonly used communication technologies in IoV systems are Cellular (2G, 3G, 4G etc.), Wi-Max, Wi-Fi, DSRC and some other 6LowPAN [12].

As illustrated in traditional RSU architecture presented in Fig. 2, communication layer is responsible for all the wired and wireless communications between RSUs and other components of IoV systems. However, due to diverse use of technologies in RSU communication, it is hard to define ethical rules for communication technologies since they operate on different frequencies in various parts of the world and are handled by different service providers. Each service provider (vendor) has its own rules for use of technology, security, privacy and data accessibility, which brings complication in the process of defining ethical guidelines of communication technologies. For example, a 6LowPAN technology, ZigBee (IEEE 802.15.4) operates on different frequencies throughout the world, e.g., 868 MHz in Europe, 915MHz in Americas and 2.4GHz elsewhere. ZigBee is used to create an IoV network to ensure cost effectiveness and less power consumption, however, due to its operation on different frequencies and multiple vendors, it becomes immensely difficult to establish ethical rules for communication that apply for all the vendors, across different operating frequencies throughout the world.

*D. Diverse Sensors*

IoV systems encompass the use of diverse sensors, e.g., speed cameras, motion detection, temperature sensors, congestion detection, vehicle detection, traffic signal monitoring and pedestrian detection etc. Each of these sensors installed in RSU perform different operations based on the context [13]. Different sensors need different rules for their operation. For example, ethical rules defined for motion detection sensors might not work for temperature sensors. Similarly, besides defining ethical rules for sensors based on their operation, sometimes, ethical rules might differ for the same sensor in different context. For example, if a vehicle detection sensor installed on a busy road has an ethical rule of capturing only the number plate of the vehicle, in normal scenarios, the rule works fine, however a fugitive with a stolen vehicle might not be captured on the road with this ethical rule and might require capturing a snapshot of driver along with the number plate as well. Diverse sensors with different context, distinct road conditions, discrete vehicles and drivers would result in substantially enormous number of scenarios each requiring an ethical rule. Such situations pose momentous challenge of stipulating ethical rules for RSUs.

*E. Dynamicity*

IoV systems are highly dynamic in nature as the topology of the network is changing rapidly due to fast moving vehicles. RSUs must be efficient and resourceful enough to respond to the requests of the vehicles in their vicinity [14]. Due to vibrant change in the network nodes, the whole network needs to be revised instantaneously to avoid delays in information transmission. In order for the requests to be acknowledged by RSUs, high processing, computing and communication is obligatory as a small delay in the information might result in serious inconvenience for the vehicles on road. For example, if a vehicle requests a RSU for traffic status on upcoming junction and RSU takes a while to gather, process and communicate this information to the vehicle; the information might not be of any use to the vehicle because by that time vehicle might reach the junction. Dynamicity poses a consequential challenge in laying the ethical rules for RSUs as the number of nodes, their locations and context changes expeditiously. The ethical rules set for RSUs based on context, number of nodes, location and type of vehicles might require processing before applying on a fleet of vehicles near the concerned RSU. By the time, the processing is done, fleet of vehicles might change because of their high speed and dynamic nature that results in waste of processing time and resources.

*F. Security and Privacy*

Security and privacy play crucial role in design and deployment of IoV systems. Due to open environment, IoV systems are exposed to several security attacks like eavesdropping, masquerading, social engineering and Denial of Service etc. These attacks can result in breach of privacy, modification of data and denial of services provided by the systems. RSUs being essential units of the IoV systems hold a lot of information including personal details of drivers and passengers. A breach in the security can seriously affect the way rules are implemented on RSUs, vehicles, drivers and passengers by law enforcing agencies [15]. Similarly, ethical rules implemented through RSUs can be modified through an attack on the system. For example, if an ethical rule of reading vehicles' number plates to calculate the toll is implemented through an RSUs and a network attacker modifies this ethical rule to take the pictures of drivers and start sending them to his server; a serious breach of driver privacy will occur that might result in unfortunate circumstances.

*G. Extensive Applications*

Employment of Internet in Vehicular Systems has unleashed voluminous opportunities of developing applications for IoVs. Vehicle manufacturers, third party RSU vendors and governmental agencies have already developed several safety and non-safety applications for VANETs and IoVs, e.g., Toll collection, navigation apps and law enforcement applications. Diverse applications for IoV are improving the experience on roads, however, encompasses a challenge of setting ethical rules for app developers. Due to varied number of developers from governmental and non-governmental sector, an ethical framework for app development and deployment specifically for RSUs is lacking that incorporates the challenge of bringing all the developer under the same ethical framework.

## IV. Proposed Ethical Principles for RSUs in IoV

Based on the review of literature, a clear gap in defining the ethical rules or framework for RSUs has been identified. Furthermore, it is evident from the literature that several challenges are faced by the ethicist to define clear rules for infrastructures (RSUs) due to dynamic nature of IoV systems, lack of architectural designs for RSUs, decentralization and scalability in IoVs, communication technologies used by RSUs to transmit information, use of diverse sensors and exposure to security and privacy attacks in IoVs. This section provides four general ethical rules for RSUs in IoVs, that are expected to set foundation for designing detailed ethical rules for different scenarios:

- Information to and from RSUs can only be monitored, modified and updated by law enforcing agencies.
- Third party hardware (sensors etc.) and software (applications etc.) for RSUs should be licensed by law enforcing agencies.
- Information of any component of IoV (vehicles, drivers, passengers and pedestrians etc.) collected through RSUs should be clearly announced before collection.
- Procedures for processing, computing, communication and information collection should be regularly monitored by law enforcing agencies.

## V. Conclusion

IoV systems are emerging in the realm of VANETs by providing additional functionality of Internet connectivity to traditional VANETs. Components of IoV are connected directly or indirectly to the Internet. The components of IoV that are not directly connected to Internet require infrastructures like RSUs to assist them in Internet connectivity. All time connected environment of RSUs brings a concern of machine ethics. This article highlighted the importance of RSUs by categorizing them into Data Collection RSUs, Toll Collection RSUs, Information Dissemination RSUs and Special Services RSUs. Considering the high importance of RSUs in IoV systems, the article focused on underpinning the challenges involved in designing the ethical rules for RSUs. Several challenges like dynamicity, lack of architectural design, decentralization and scalability, diversity of sensors, extensive applications and security and privacy have been highlighted to set a solid foundation of proposing general ethical rules for RSUs in IoV systems. Finally, the article proposed four general ethical rules for RSUs that are expected to lay a strong structure for designing ethical principles for different layers of RSU architecture.